\documentclass[letterpaper, 10 pt, conference]{ieeeconf}  % Comment this line out
\IEEEoverridecommandlockouts                              % This command is only
\usepackage{graphicx}

\usepackage[hyphens]{url}
\usepackage[hidelinks]{hyperref}
\hypersetup{breaklinks=true}
\usepackage{cite}

\let\labelindent\relax
\usepackage[inline,shortlabels]{enumitem}
\setlist{noitemsep,leftmargin=*}
\usepackage[colorinlistoftodos,prependcaption,textsize=tiny]{todonotes}

\title{%\LARGE \bf
AIOps for a Cloud Object Storage Service%\newline
}

\author{%
	Anna Levin, Shelly Garion, Elliot K. Kolodner, \\
	Dean H. Lorenz, Katherine Barabash\\
	IBM Research -- Haifa\\
	\texttt{\small \{lanna,shelly,kolodner,dean,kathy\}@il.ibm.com}
	\and
	Mike Kugler, Niall McShane\\~\\
	IBM Cloud and Cognitive Software\\
	\texttt{\small Mike.Kugler@ibm.com, nmcshane@us.ibm.com}
}

\begin{document}

\begin{titlepage} 

\begin{center}
\vspace*{5cm}

\textbf{\LARGE{ AIOps for a Cloud Object Storage Service}}

\vspace*{1cm}
Anna Levin, Shelly Garion, Elliot K. Kolodner, Dean H. Lorenz, Katherine Barabash
IBM Research – Haifa
lanna,shelly,kolodner,dean,kathy@il.ibm.com

\vspace*{0.5cm}
Mike Kugler, Niall McShane
IBM Cloud and Cognitive Software
Mike.Kugler@ibm.com, nmcshane@us.ibm.com

\vspace*{1cm}
Published in: 2019 IEEE International Congress on Big Data (BigDataCongress)

\vspace*{1cm}
$\copyright$2019 IEEE. Personal use of this material is permitted. Permission from
IEEE must be obtained for all other uses, in any current or future media,
including reprinting/republishing this material for advertising or promotional
purposes, creating new collective works, for resale or redistribution to servers
or lists, or reuse of any copyrighted component of this work in other works.
DOI: 10.1109/BigDataCongress.2019.00036

\vspace*{1cm}
This work has been partially supported by the SODALITE project, grant agreement 825480, funded by the EU Horizon 2020 Programm.

\end{center}
\end{titlepage}

\maketitle
\thispagestyle{empty}
\pagestyle{empty}

%%%%%%%%%%%%%%%%%%%%%%%%%%%%%%%%%%%%%%%%%%%%%%%%%%%%%%%%%%%%%%%%%%%%%%%%%%%%%%%%
\begin{abstract}

With the growing reliance on the ubiquitous availability of IT systems and services, these systems become more global, scaled, and complex to operate. 
To maintain business viability, IT service providers must put in place reliable and cost efficient operations support.
Artificial Intelligence for IT Operations (AIOps) is a promising technology for alleviating operational complexity of IT systems and services. 
AIOps platforms utilize big data, machine learning and other advanced analytics technologies to enhance IT operations with proactive actionable dynamic insight. 

In this paper we share our experience applying the AIOps approach to a production cloud object storage service to get actionable insights into system's behavior and health. 
We describe a real-life production cloud scale service and its operational data, present the AIOps platform we have created, and show how it has helped us resolving operational pain points.

\end{abstract}

%%%%%%%%%%%%%%%%%%%%%%%%%%%%%%%%%%%%%%%%%%%%%%%%%%%%%%%%%%%%%%%%%%%%%%%%%%%%%%%%
\section{Introduction} \label{sec:intro}

Information Technology (IT) has transformed almost all industries and areas of human life. Technology has gone all the way from automating tedious computations and eliminating paper-driven office processes to governing life-saving surgery and competing with humans on trivia, game, and debate contests. As humanity's reliance on computing becomes ubiquitous, IT installations grow larger and more complex, demanding increasingly more resources for their bring up and operation. Cloud-scale IT operators own tens of data centers full of compute, network, and storage devices, running complex multi-layer software stacks and hosting a multitude of clients. Cloud-scale service providers develop and run multi-region multi-datacenter solutions under ever increasing availability, performance, and security demands. Till very recently, IT operators, while offering advanced data-driven analytics to their clients in various domains, employed old-fashioned manual processes for running their own business operations.  Finally, it has become apparent that to cope with growing operational complexity and costs, the IT business itself requires digital transformation.

To highlight the growing interest and investment in this transformation, Gartner has introduced the concept of AIOps~\cite{GartnerAIops}. 
The concept originally stood for Algorithmic IT Operations and later became known as Artificial Intelligence for IT Operations. Today, most planet-scale service operators employ their own AIOps to collect logs, traces and telemetry data, and analyze the collected data to enhance their offerings~\cite{dapper,ebay,facebook,netflix}. In addition, multiple new vendor products for smaller scale operators and enterprise IT have been created or rebranded as AIOps~\cite{Moogsoft,OpsRamp,Zenoss}. Some AIOps platforms are intrinsically integrated into the IT system they help to operate. Other solutions, mostly vendor products and services, are more generic. These often cover only a part of the AIOps stack, such as data collection, or specific analytical methods and algorithms, or data integration in a data lake, etc. Such point solutions are hard to integrate into existing operational production systems in a meaningful way. We have addressed this gap by creating a useful AIOps system around an existing cloud-scale service, namely 
an
%a cloud-scale 
object storage service. Cloud-scale object storage may contain trillions of objects, serve massive amounts of simultaneous access requests, and generate tons of operational data. Our  goal is extracting useful business insights from the operational data collected dynamically from a production service. However, the sheer magnitude of data, as well as variability of formats and data sources, and speed of data generation, make it difficult to turn abstract theories into practice. 
%AIOps tools and platforms are data driven and evolve around data collection, data storage, data analytics, and data visualization. Combined, these data technologies form a notion of a data lake, a next-generation data storage and management solution to meet the ever-evolving needs of increasing volumes of big data.

In this paper, we share our experience and lessons learned creating a data driven AIOps platform for a production cloud object storage service at IBM. Our contributions are: (1) describing service operations in production and the available operational data, (2) presenting the AIOps solution we have built for gaining operational insights, and (3) showing the operational pain points our solution helps to resolve. %The paper is structured as follows: Section~\ref{sec:problem} provides the problem definition along with our goals and requirements; Section~\ref{sec:data} covers the data processing pipeline; Section~\ref{sec:analysis} describes  the analytics and the delivery of AIOps results; Section~\ref{sec:concl} concludes the paper, summarizing the lessons learned and outlining future research plans.

\section{Problem Definition}
\label{sec:problem}
We are developing AIOps capabilities for IBM Cloud Object Storage 
(COS)
so
that its operation will be data driven and automated.
We collect several types of operational data from object storage,
and do several different analyses on it.
%First we briefly describe IBM Cloud Object Storage,
%then the relevant data sources that we are collecting,
%and finally our overall goals and the challenges that we faced.
%
%\subsection{IBM Cloud Object Storage}
%IBM Cloud Object Storage
%encrypts and disperses the data stored in it 
IBM COS encrypts and disperses the objects stored in it
using erasure coding
across multiple geographic locations~\cite{resch_2011}.
Access to objects is over HTTP
using a REST API.
%IBM Cloud Object Storage is offered with three degrees of resiliency:
%(1) Cross Region, where the data is dispersed across three data
%centers (availability zones) in three different regions in the same geography,
%(2) Regional, where the data is dispersed across three data centers
%(availability zones) in the same region, and
%(3) Single Site, where the data is dispersed across multiple servers
%within the same data center.
%\begin{figure}
%\centering
%\includegraphics[width=0.5\textwidth]{Figures/worldwideCOSmap.png}
%\caption{IBM Cloud Object Storage offerings}
%\label{fig:worldwideCOSmap}
%\end{figure}
%Figure~\ref{fig:worldwideCOSmap} shows the worldwide offerings.
IBM COS has a two tier architecture:
(1) front-end servers, called {\em Accesser\textsuperscript{\textregistered}} nodes,
that  receive  the  REST  commands  and  then  orchestrate  their
execution across
(2) storage-rich back-end servers, called {\em Slicestor\textsuperscript{\textregistered}} nodes, that store
the data.
%Accessers communicate with Slicestors to store and retrieve data;
%Slicestors communicate with other Slicestors to recover lost data,
%e.g., when a disk fails,
%and to redistribute data,
%e.g., when additional Slicestors are added to a system.

\subsubsection{Operational data}
There are many types of 
IBM COS
operational data, e.g., logs and metrics.
%, available from an IBM COS system.
We describe two kinds that we use in our analyses.

The first are access logs~\cite{access_logs}.
These are in JSON format.
These logs contain an entry for each operation invoked on the object storage.
The entry contains a wealth of information regarding the operation;
this includes items such as the operation type, the bucket and object
names on which the operation works, the HTTP return code, the start
and end times of the operation, and various latency statistics (total
latency, time spent waiting for the client, time spent waiting for the
storage back end, time for authentication and authorization, etc.).
The access logs are generated on the Accesser nodes.

The second are connectivity logs.
These are structured 
JSON 
logs %in JSON format 
produced once a
minute on each
IBM COS server.
% implementing IBM COS.
They provide information about the connectivity of the servers across all the offerings of IBM
COS.
%In particular, the entries from each server include two lists:
%(1) a list of other servers that the application on this server
%succeeded in communicating with and the average application-level
%latency for communication with that server over the last minute, and
%(2) a list of the other servers with that application could
%not establish communication with over the minute.

\subsubsection{Goals and challenges}
Our aim is to detect, predict and prevent failures and performance slowdowns
that could impact users.
Cloud object storage is engineered with redundancy,
so that it continues working despite failures of individual components.
Thus, when a failure occurs
it is masked by the redundancy% 
, which 
% and this 
makes it difficult for the operators to discover.
Our goal is to discover these failures and pinpoint their cause
through %the 
analysis of the %available operations
 data.

In turn, the analysis of the operations data also poses several challenges.
The first is that the schema of the log data is dynamic.
Each log record is JSON, so its fields are labeled, but each
individual record might have a different subset of the fields, e.g.,
depending on the operation type,
the authentication mechanism,
or whether the object data is encrypted.
The second is the scale of the data to be processed.
As described in Section~\ref{sec:curation},
we converted the JSON data to Apache Parquet,
partitioned it and added derived fields
in order to save space and speed subsequent processing.
Poor choices, e.g., for partitioning, or producing Parquet objects that are too small,
could lead to the costly need to reconvert and reprocess the data.
Also, programs tested on small samples of data
might fail after long running times when processing very large amounts of data,
e.g., when encountering a new log record instance.

\subsubsection{Our Approach}\label{sec:background}
%The high level description of our flow from gathering through the data lake consists of the following stages:

%In particular, we keep the data in full in a Cloud Object Storage~\cite{COS} and analyze it using Apache Spark~\cite{ApacheSpark},~\cite{zaharia_2016}. This approach leverages Netflixâs %Atlas~\cite{netflix} tool, in which the data at full resolution is kept in Amazon S3~\cite{amazonS3}, and Apache Hadoop~\cite{hadoop} (Elastic MapReduce~\cite{EMR}) is used for processing %the data.

%Our flow consists of the following stages:
%\begin{figure}
%\centering
%\includegraphics[width=0.5\textwidth]{Figures/DataFlow2Lake_v4.PNG}
%\caption{Data analytics pipeline}
%\label{fig:flow2lake}
%\end{figure}
\begin{figure*}[!ht]
	\begin{minipage}[t]{0.5\linewidth}
		%\begin{figure}
		\centering
		%includegraphics[width=0.48\textwidth]{Figures/Flow_HDFS_v2.png}
		\includegraphics[width=0.9\textwidth]{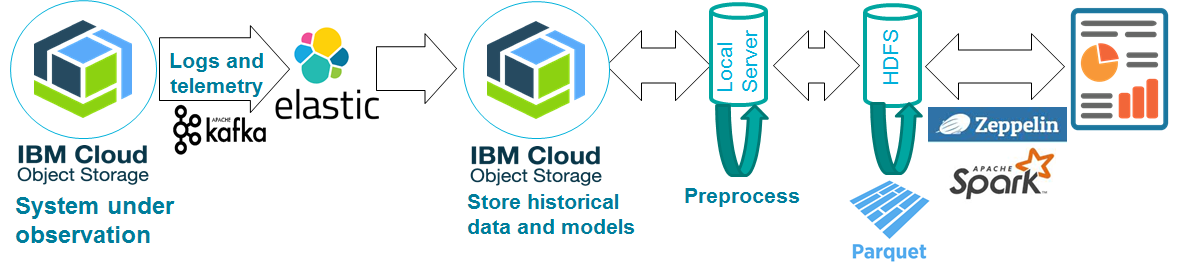}
		\caption{Pipeline for processing operational logs}
		\label{fig:local_flow}
		%\end{figure}
	\end{minipage}
	\hfill
	\begin{minipage}[t]{0.5\linewidth}
		%\begin{figure}
		\centering
		\includegraphics[width=0.9\textwidth]{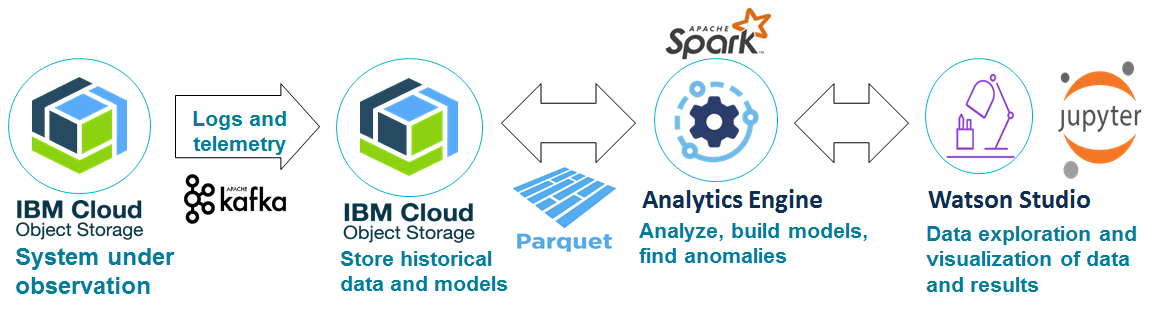}
		\caption{Cloud-based pipeline}
		\label{fig:cloud_flow}
		%\end{figure}
	\end{minipage}
	\vspace{-2ex}
\end{figure*}

%Our approach to AIOps includes the following stages:
We apply the following stages:
\begin{description}
\item[Ingestion (Sec.~\ref{sec:injestion}):]
Ingestion of historical data is from batch files, e.g., Elasticsearch dump files,
as well as of streaming data, e.g., from Apache Kafka.

\item[Curation (Sec.~\ref{sec:curation}):] 
Data curation, cleaning and preparation for analytics is crucial, and requires a huge investment. 
In data warehousing this stage is called Extract, Load and Transfer (ELT).

\item[Features (Sec.~\ref{sec:features}):] 
Features need to be generated in order to use statistical and machine learning algorithms.
%In order to use statistical and machine learning algorithms, it is required to generate the features for the analytics. 
We present a ''smart groupBy'' method to generate  features efficiently in parallel at scale using Apache Spark.

\item[Model (Sec.~\ref{sec:model}):]
%Both statistical approach and machine learning methodologies were used for performing the desired analysis.
We use both statistics and machine learning to create our models and analyze incoming data.

\item[Causality (Sec.~\ref{sec:cause}):]
%After performing the analysis, 
We do feature isolation to detect the root cause of a problem, such as a failing component. 

\item[Visuals (Sec.~\ref{sec:visual}):]
%Finally, the root cause of the problem should be presented to the operator in order to take action.
Finally we present the root cause of the problem to the operator through
%This can be done in
reports, dashboards (e.g., Grafana), 
or by notifications (e.g., Slack).
\end{description}

%%%%
%\emph{Explain our solution and mention alternatives and reasons behind the selected technologies}.
%%%

\section{Data Processing} \label{sec:data}
Our data processing flow was implemented in two ways.
Initially the data samples were copied to a local storage system for easier exploration, see Fig.~\ref{fig:local_flow}.
However, moving the data far from its collection became unfeasible for the real logs sizes we faced.
Thus, we now do the data processing flow in the cloud, as shown in Fig.~\ref{fig:cloud_flow}.

Fig.~\ref{fig:local_flow} shows the initial data processing flow.  The IBM COS access logs in JSON format are collected in Elasticsearch~\cite{elastic} (ES) and historical dumps of ES are stored in an IBM COS bucket.
In order to build a non-intrusive exploration pipeline, we copied the ES dumps to a local storage system:  splitting files into chunks to optimize parallel copying and verifying correctness with MD5.
The copied chunks were stored in HDFS~\cite{hadoop}
%(Hadoop Distributed File System, see~\cite{hadoop})
and converted to Parquet~\cite{parquet} to ease the analytics.
%, as described in Sec.~\ref{sec:curation}.
For analytics we used %Apache
Spark together with Zeppelin notebooks~\cite{zeppelin}, a common platform for interactive data analytics.
%\begin{figure*}
%\begin{minipage}[t]{0.5\linewidth}
%%\begin{figure}
%\centering
%%includegraphics[width=0.48\textwidth]{Figures/Flow_HDFS_v2.png}
%\includegraphics[width=0.9\textwidth]{Figures/Flow_HDFS_v2.png}
%\caption{Pipeline for processing operational logs}
%\label{fig:local_flow}
%%\end{figure}
%\end{minipage}
%\hfill
%\begin{minipage}[t]{0.5\linewidth}
%%\begin{figure}
%\centering
%%\includegraphics[width=0.48\textwidth]{Figures/Flow_cloud_v2.png}
%\includegraphics[width=0.9\textwidth]{Figures/Flow_cloud_v2.png}
%\caption{Cloud-based pipeline}
%\label{fig:cloud_flow}
%%\end{figure}
%\end{minipage}
%\vspace{-2ex}
%\end{figure*}

Despite the advantages of having data close to the data scientist for efficient development and easy data manipulation, the initial solution was not scalable enough for our needs.
As a result, we now do the processing in the cloud, as depicted in Fig.~\ref{fig:cloud_flow}.
In this second pipeline, the data processing is integrated with the collection stage and the logs are stored in IBM COS in Parquet format.
The data is consumed for analytics by Apache Spark~\cite{spark} applications running on the IBM Analytics Engine~\cite{IAE}. %, as described in Sec. \ref{sec:analysis}.
For interactive development, we use Jupyter notebooks~\cite{Jupyter} provided by IBM Watson Studio~\cite{DSX}.
In the remainder of this section we describe our experience with the processing flows presented above.

%\begin{figure}
%      \centering
%    \includegraphics[width=0.48\textwidth]{Figures/Flow_cloud_v2.png}
%    \caption{Cloud-based pipeline}
%     \label{fig:cloud_flow}
%\end{figure}

\subsection{Ingestion}\label{sec:injestion}
%Placement options (close to processing, close to creation, \emph{object store} vs DB/ElasticSearch)
%Format (logstash/Elastic snapshots/\bf{stream})
One of the common ways to ingest log files into a data lake is a Logstash processing pipeline~\cite{logstash} that ingests data from multiple sources simultaneously, transforms it, and then sends it to Elasticsearch - an open-source, RESTful, distributed search engine.
We started with this ingestion approach in Fig~\ref{fig:local_flow}.
IBM COS operational logs were collected from the Accesser nodes, sent over Apache Kafka and stored in ES through Logstash.
The ability to test development directly against ES before working with the full data volume provided fast initial time to value because of the easy integration with Spark and with visualization and reporting tools, e.g. Kibana.
%The additional advantage of ingesting data with ES is its snapshot mechanism used for backup and retention, and allowing non-intrusive data exploration. For backup we took incremental snapshots from a running ES cluster and stored them in a remote repository using IBM COS plugin.To start initial exploration of the data without intervention with production systems, we restored the logs from IBM COS to the local servers and then ingested them into HDFS, as shown in Fig.~\ref{fig:local_flow}.

We started the development of the pipeline by ingesting the access logs through ES to a local HDFS system, as depicted in Fig.~\ref{fig:local_flow}.
The log data was moved to HDFS, close to the processing, and after cleaning and format adjustment, the data was consumed for analytics. 
This method of ingestion provided an isolated environment for exploration independent from the production pipeline.
%It allowed us to ingest the data once and work with it on our own resources within our own sandbox, without being affected by changes (e.g., schema updates).
This approach worked well for exploring relatively small samples; however, as the volume of the log data grew,
%and crossed hundreds of GB per day,
it became infeasible to move the logs to HDFS.

Therefore, we decided to switch to a pipeline based on cloud services shown in Fig.~\ref{fig:cloud_flow}.
In this approach we collect and stream the logs directly to IBM COS, storing them close to the collection points and organizing them in Parquet, and also move the processing close to the data creation.
Note that while this cloud-based approach is closer to a full data-lake pipeline, it is still a development environment.
%Rather than catalogue the raw data and provide a data warehouse view, we created our own structured copy as a data source.
This approach is useful during development as it provides a sandbox with a "cached" snapshot of the data.
As demonstrated in the next section, this approach allows exploration of analytic methods in a repeatable manner without having to re-run the entire data pipeline.

\subsection{Curation}\label{sec:curation}
It is not enough to collect the data to make it useful and meaningful.
The ingested data must be cleaned, formatted and organized in order to make it useful for data scientists engaged in data discovery and analysis.
The process of managing data, archiving and representing is called data curation.
Data curators collect data from diverse sources and integrate it into repositories that are many times more valuable than the independent parts.
Curation also ensures data quality and makes machine learning (ML) more effective.

It is hard to overestimate the importance of data cleaning and sanity checking.
Our sanity checks include row counts, timestamps consistency checks and statistical data validation.
%In a huge data set, it is expected that the number of rows preserves its order of magnitude.So we started with a simple row count and when there was a significant change in the number of rows, we understood that there is a problem in data collection process.The decrease in the number of rows means data is lost and not written properly, the increase may mean the data was written twice or more.
In addition to counts, it is important to validate that statistical characteristics are preserved, e.g., the mean and median values of the important numerical fields.
Further data cleaning and standardization require a unifying format for numerical data to allow numerical operations, and filling in missing data with appropriate values, e.g., null values of the appropriate type.
Another important aspect of data cleaning is validating timestamps consistency and correctness, needed to ensure that no data is missing, ignored or duplicated.
Validating timestamp consistency is a complex task in a heterogeneous system.
Our data set is collected from different systems using multiple formats across many timezones.
In order to validate timestamp consistency and enable data analysis, we unified the format and converted all timestamps to UTC.
%There are multiple helpful libraries available in python, scala and other languages, which ease the process of timestamps conversions.

In the data lake pipeline, curation prepares the data for consumption by analytic tools.  Curation transforms the raw data into a structured view, annotates it using metadata, combines current inputs with historical data, and integrates previously aggregated data.
During development, we chose to use a staging transformation approach.
Staging the data in Parquet format greatly speeds up the debugging process while developing new analytic methods; it reduces the time wasted waiting for computations to complete.  Moreover, staging simplifies repeatable experiments.
We experimented with several choices before deciding on the right staging transformations and format to achieve the best performance.

Apache Parquet is a columnar data format, which supports compression and encoding schemes.
The schema is embedded in the data itself, so it is a self-describing format.
These features make it efficient to store data in Parquet as opposed to row-oriented schemes like CSV and TSV.
Fig.~\ref{fig:parquet_size} shows the results of our experiments on the effectiveness of storing our log samples in Parquet.
The figure shows for our case, when no explicit options are set and Spark uses default snappy compression, storing logs in Parquet rather than leaving them in JSON format in a form of compressed ES indexes, reduced the size by more than 10 times.
%\begin{figure}
%      \centering
%    \includegraphics[width=0.48\textwidth]{Figures/ParquetSize.png}
%    \caption{Size ratio Parquet vs. Raw JSON Logs Index Compressed}
%     \label{fig:parquet_size}\vspace{-2ex}
%\end{figure}

In addition, the columnar nature of Parquet enables efficient querying since most queries involve few columns and can be enhanced further by partitioning based on the values of one or more columns.
It is crucial to know the common queries in order to benefit from by data partitioning.
For example, among other fields our schema involves dates, locations and customer information.
Since our typical analysis was done on a daily basis, we decided to partition data by date and location, storing it in an order dictated by dates and location columns.
When the query is done for specific day and location only small sub-set of data is accessed.
This order of partitioning served us well for daily performance logs analysis, but proved to be inefficient when per customer analysis was requested.
In order to access specific customer data, we needed to access whole data set regardless of its date and location, losing all the advantages of partitioning.

Partitioning not only has benefits for performance, it also helps managing the data.  Spark supports programmatic partitioning and can discover partitions automatically if the stored data is already partitioned.
To store partitioned data in the most efficient way we performed experiments with Spark to determine the best way to write the data.
The following write options were evaluated:
\begin{enumerate*}[(1)]
    \item Each day-location gets its own dataset.
    \item Hive-style partitioning using partition\_by.write in \emph{append} mode.
    \item Hive-style partitioning explicitly writing each virtual directory in \emph{overwrite} mode.
\end{enumerate*}
Our experiments showed that option 1 is the slowest one. Options 2 and 3 are similar in terms of performance and twice as fast as option 1. The latter two options store the data and access it in a similar way;
however, there is a difference between options 2 and 3 when a write job fails. 
If Spark job fails in option 2 in the middle of data writing, it cannot be easily reverted as the written data becomes part of a big data set with a single flag for successful writes. 
As a result retrying the write job might create duplicate data chunks. On the other hand, with option 3, if the write job fails writing a virtual directory, the partial results can be easily cleaned by rerunning the job, overwriting the specific directory.
Therefore, we chose hive-style partitioning, explicitly writing a virtual directories in \emph{overwrite} mode, as the fastest and safest option.

\subsection{Feature extraction}\label{sec:features}
%Analytic process considerations:
%\begin{itemize}
%\item Spark vs just python
%\item Notebooks vs spark submit
%\end{itemize}

In large scale data sets, such as operational log data (e.g.~\cite{AWSlogs}), a big challenge is preparing the raw data for use with statistical and ML methods. In the language of ML the prepared data is called ''features''.
Moreover, for many statistical and ML methods, such as anomaly detection, choosing and generating the right features is much more important than the actual algorithm. For example, with the proper features basic outlier detection methods can find anomalies (see the discussion in~\cite{ebay}).

We use Spark~\cite{spark, ApacheSpark}, which supports the Map/Reduce programming model,
to generate the features efficiently in one parallel pass over the huge amount of log data.
For the development of algorithms and code we use Jupyter notebooks, and for running the code in batch in production we
use "spark-submit" to the IBM Analytics Engine~\cite{IAE}.
We describe our ''smart groupBy'' method to generate the features efficiently,
assuming that the input data is in a table format (e.g., Spark DataFrame) and that the processing is done using an SQL API (e.g., Spark-SQL).
\begin{description}
\item[Map Step] We enrich the input data 
with additional 
%set with the following 
columns:
%    \begin{description}[labelsep=0,format={\normalfont\emph}]
        %\item[Ranges and bucketing:] 

        \emph{1) Computing ranges and buckets}:
        by ranges of timestamps (e.g., day/hour/minute), %ranges of 
        sizes, %as well as
        and 
        other numerical components.
        %\item[Reducing a large set of items to a small subset (usually referred as Ã¢ÂÂdata cleaningÃ¢ÂÂ):]

        \emph{2) Reducing a large set of values to a smaller subset (often called data cleaning)}:
        by parsing the original values and taking only dominant substrings, or taking only the most popular subset of values, etc.
        %\item [Computing a function of several columns:]

        \emph{3) Computing a function over several columns}: 
        by choosing a derived value based on values appearing in the input columns.
        %\item []Derivations and differences:]

        \emph{4) Doing derivations and differences}:
        by computing a derived column $C$, adding a new column $C_{shift}$, which is a copy of $C$ shifted in one cell, and then computing the derived column $C_{derived} = C-C_{shift}$.
        %\item [Statistical functions:] 

        \emph{5) Computing statistical functions}:
        counts, mean, standard deviation, ranks, percentiles, etc.
%    \end{description*}

The one pass pipeline combines these transformations into a single map step.
\item[Reduce Step]
    We perform one SQL ''group by'' operation on many columns at once, including the columns generated in the map step, to produce all the features in parallel.
\end{description}

In order to choose the specific ranges, buckets, and subsets in steps 1 and 2, we use prior domain knowledge or learn from a small sample of the large data set.
The statistical functions in step 5 may also be used to check the reliability of the generated features, for example, checking whether the number of items per feature is statistically meaningful.
Since Spark is lazy~\cite{spark}, the entire calculation is done on-demand in the ''Reduce step'' to obtain the features. We perform the ML algorithm or statistical analysis on these features.

\section{Analytics and Insights}\label{sec:analysis}
In order to gain insight we perform statistical analysis and create ML models using the extracted features.
Our analysis is done using Python packages, such as pandas~\cite{pandas} for general statistics,
and scikitlearn~\cite{scikitlearn} for ML.

\begin{figure*}
	%	\begin{table}
	\begin{minipage}[t]{0.23\linewidth}
	%\begin{figure}
		\centering
		\includegraphics[width=\linewidth]{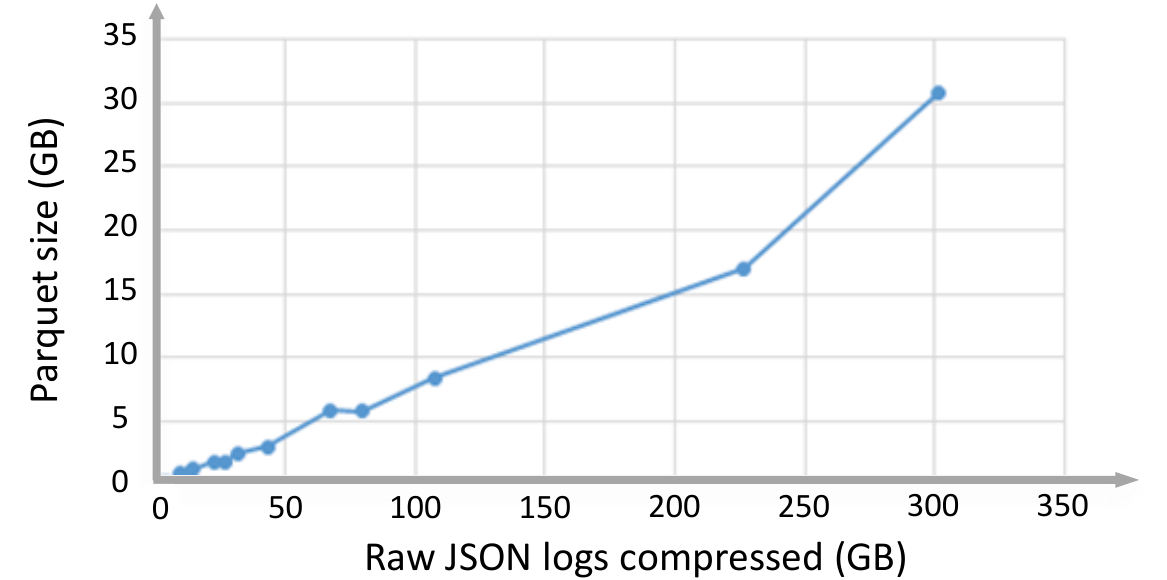}
		\caption{Size ratio Parquet vs. Raw JSON Logs Index Compressed}
		\label{fig:parquet_size}\vspace{-2ex}
	%\end{figure}
	\end{minipage}
\hfill
	\begin{minipage}[t]{0.23\linewidth}
		%\begin{figure}
		%\centering
%		\includegraphics[width=\linewidth,height=12ex]{Figures/AnomalyScore.PNG}
		\includegraphics[width=\linewidth]{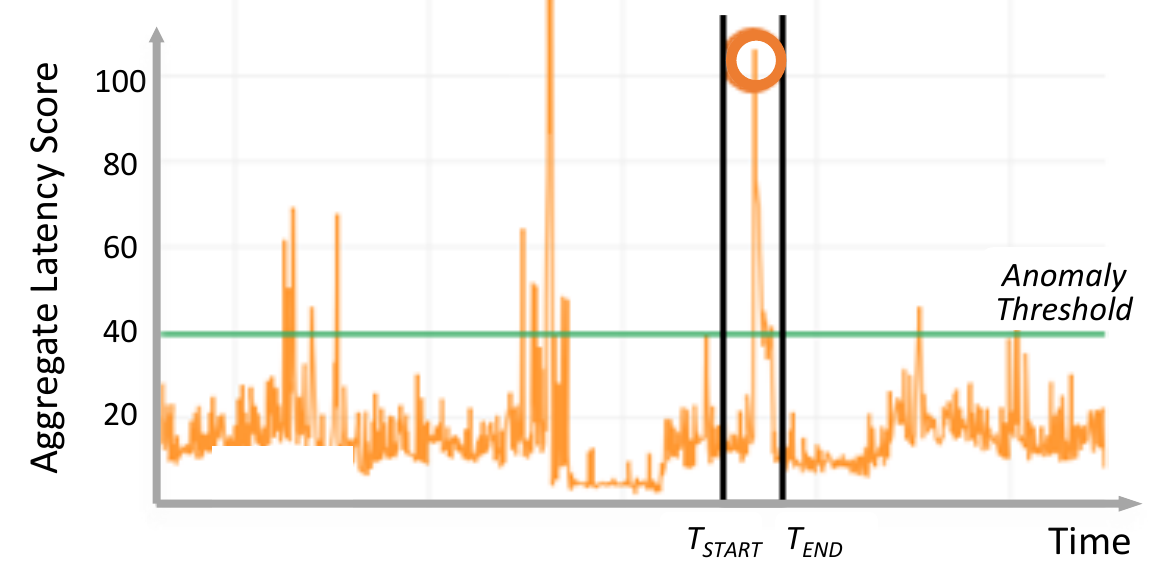}
		\caption{Anomaly score of the aggregated latency}
		\label{fig:AnomalyScore}
		%\end{figure}
	\end{minipage}
\hfill	%&
	\begin{minipage}[t]{0.23\linewidth}
		%\begin{figure}
		%\centering
%		\includegraphics[width=\linewidth,height=12ex]{Figures/Anomaly_Latency_v2.PNG}
		\includegraphics[width=\linewidth]{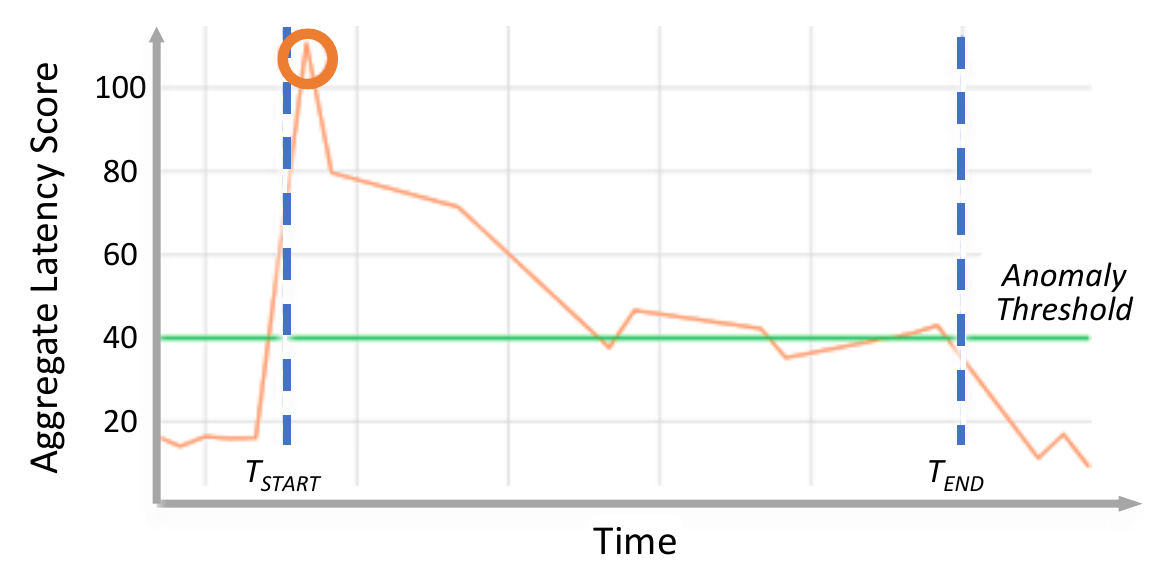}
		\caption{Anomaly score of the aggregated latency --
%			focus on the anomaly}
			zoom in}
		\label{fig:AnomalyLatency}
		%\end{figure}
	\end{minipage}
\hfill	%&
	%%%%
	%{\bf Add another graph with label: fig:ConnectivityTool }
	\begin{minipage}[t]{0.23\linewidth}
		%\begin{figure}
		%\centering
%		\includegraphics[width=\linewidth,height=12ex]{Figures/ConnectivityIssue.png}
		\includegraphics[width=\linewidth]{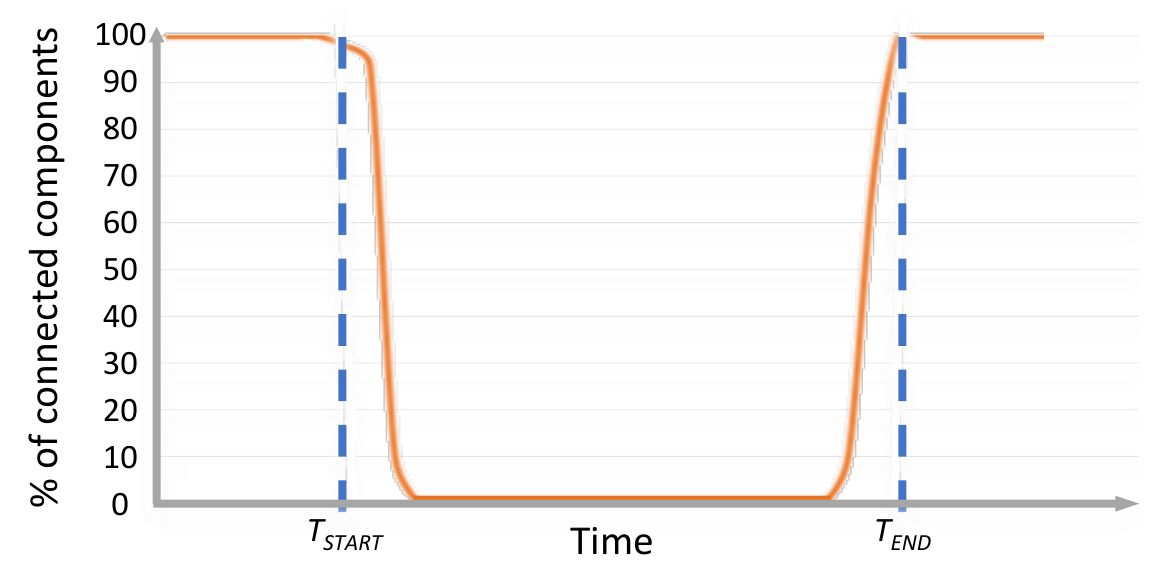}
		\caption{Connectivity point of view: disconnected component}
		\label{fig:ConnectivityTool}
		%\end{figure}
	\end{minipage}
	\vspace{-3ex}
	%\end{table}
	
\end{figure*}
\subsection{Analytical model: statistical or machine learning (ML)}
\label{sec:model}
%\emph{Add paragraph on each approach in general with pros and cons. Our selection reasoning and graph -- %Simple statistics/aggregated metrics/anomaly detection}

%There are several methodologies that we have used to perform the desired analysis.
In our use-cases we are able to provide an efficient solution using basic statistical and ML methods, and did not need deep-learning methods, which are less efficient and require more resources, especially when implemented at large scale.

\emph{Statistical approach.} Computing statistical invariants of the large-scale dataset, such as counts, mean, standard deviation, histograms, median and other percentiles, is extremely useful to provide an overview of the operations data and perform basic reliability and sanity checks. This calculation can also be done efficiently at large scale using the ''smart groupBy" method described above.
We perform hierarchical aggregation of various metrics in order to detect and focus on a failed component, as can be seen in
Fig.~\ref{fig:ConnectivityTool}, to find the highest possible level of aggregation/hierarchy,
representing the most significant problem currently occurring in the system.
However, in cases where there are too many metrics and signals, it is not possible to manually handle too many graphs and alerts, 
so
%and 
it is better to use %an 
automated ML tools.

\emph{Machine learning approach.}
%There are two kinds of machine learning methods, the \emph{supervised} approach, in %which one learns from labels attached to a training dataset, and the %\emph{unsupervised} approach, in which no such labels are given. Since labeling the %data requires a lot of investment, especially at large scale, 
%We have used an unsupervised \emph{anomaly detection} tool. 
%as well as \emph{clustering} methods.
There are various \emph{anomaly detection} methods, for an extensive survey see~\cite{AnomalyDetection}. The basic approach is based on calculating the \emph{z-score} of a single metric, namely, the signed fractional number of standard deviations by which the value $x$ of an observation differs from the metric's mean value.
%, that is $|x-mean| / std$. There are also more robust methods, for example, based on calculating $|x-median| / MAD$ where MAD is the Median Absolute Deviation, see~\cite{ebay} for a discussion and comparison.
In addition to a univariate approach, there exist multivariate anomaly detection methods, e.g., ~\cite{MultiAnomaly}, which reduce false alerts.
We use a %basic 
multivariate anomaly detection algorithm on the features %that we 
generate using the ''smart groupBy'' method described in Sec.~\ref{sec:features}. These features include aggregations of various latency metrics of several components in the system, and our algorithm is in the spirit of Ng's algorithm in~\cite{NgML}.
Fig.~\ref{fig:AnomalyScore} shows a graph of our aggregated anomaly score and computed threshold ; Fig.~\ref{fig:AnomalyLatency} focuses on the anomaly inside the rectangle.

%In addition, we used clustering methods in order to understand the customer %workloads and usage patterns, discover errors in the way that they are using cloud %object storage, find issues that might be common to multiple customers, and make %recommendations. We characterized each customer according to its set of most %dominant operations, and consequently, clustered similar customers according to %their distribution of operations. Then we could easily identify customers that have %an anomalous workload.

%\begin{figure*}
%%	\begin{table}
%\begin{minipage}[t]{0.3\linewidth}
%%\begin{figure}
%%\centering
%\includegraphics[width=\linewidth]{Figures/AnomalyScore.PNG}
%\caption{Anomaly score of the aggregated latency}
%\label{fig:AnomalyScore}
%%\end{figure}
%\end{minipage}
%%&
%\begin{minipage}[t]{0.3\linewidth}
%%\begin{figure}
%%\centering
%\includegraphics[width=\linewidth]{Figures/Anomaly_Latency_v2.PNG}
%\caption{Anomaly score of the aggregated latency --
%focus on the anomaly}
%\label{fig:AnomalyLatency}
%%\end{figure}
%\end{minipage}
%%&
%%%%%
%%{\bf Add another graph with label: fig:ConnectivityTool }
%\begin{minipage}[t]{0.3\linewidth}
%%\begin{figure}
%%\centering
%\includegraphics[width=\linewidth]{Figures/ConnectivityIssue.png}
%\caption{Infrastructure point of view: disconnected component}
%\label{fig:ConnectivityTool}
%%\end{figure}
%\end{minipage}
%%\end{table}
%
%\end{figure*}

\subsection{Root cause and problem isolation}\label{sec:cause}
%\emph{Add graphs correlation from hackathon
%Address all use cases
%Explain your findings in user's way/language
%Focus on a relevant component/feature
%Pinpoint to the problem
%Features isolation}
After completing the statistical and ML analysis, we perform feature isolation in order to find the root cause and detect the failed component.
Figs.~\ref{fig:AnomalyLatency} and ~\ref{fig:ConnectivityTool} demonstrate two equivalent points of view of the same failure that occurred in our system at the same time (from time $T_{start}$ to $T_{end}$).

Fig.~\ref{fig:ConnectivityTool} shows the \emph{connectivity} point of view.  We observe that certain application components were disconnected from the other components from time $T_{start}$ to $T_{end}$.
In order to detect the most relevant problematic component for an alert, we reduced the problem of determining the failing components to a hierarchical flow problem.
This approach allows us to pinpoint the problem at the most relevant level of aggregation/hierarchy, and notify the operations team of the most significant problem currently occurring in the system.

Fig.~\ref{fig:AnomalyLatency} shows the \emph{performance} view.  Our multivariate anomaly detection tool on the latency metrics shows a high peak at time $T_{start}$  that stays above the anomaly threshold for the same period from $T_{start}$ to $T_{end}$.
In this case, we isolate the problem and focus on the misbehaving component by indicating the top features with the highest z-scores during this anomalous period, and checking how many of them come from the same component.

%\begin{figure}
%\centering
%\includegraphics[width=0.5\textwidth]{Figures/Anomaly_Latency_v2.PNG}
%\caption{Anomaly score of the aggregated latency --
%focus on the anomaly}
%\label{fig:AnomalyLatency}
%\end{figure}
%
%%%%%
%%{\bf Add another graph with label: fig:ConnectivityTool }
%
%\begin{figure}
%\centering
%\includegraphics[width=0.47\textwidth]{Figures/ConnectivityIssue.png}
%\caption{Infrastructure point of view: disconnected component}
%\label{fig:ConnectivityTool}
%\end{figure}

\subsection{Visualizations and dashboards}\label{sec:visual}
%\begin{itemize}
%\item Static reports/semi-static with values when point/interactive
%\item Dashboard vs e-mail/slack/pager duty notifications  - action points
%\end{itemize}

In order to provide alerts and reports to the operation team, one can use static reporting tools that mainly present periodical graphs and textual reports.
Another option is presenting semi-static tools, which allow interactive exploration of the pre-calculated results and support drill down and zoom-in at problematic points, such as Grafana~\cite{grafana}.
Moreover, such semi-static tools allow to incorporate all the produced graphs into a few dashboards.
However, such tools do not cover all types of required insights and graphs.
For example, in our use-cases a connectivity matrix heatmap turned-out to be the most useful tool for presenting failed component in the context of overall system.

In addition to the graphical tools and dashboards, when an action of an operator is needed, it is necessary to provide a direct alert or real-time notification via tools like slack~\cite{slack}.
For example, our automated tool provides a slack notification to the operator, containing the identity of the specific application or network component experiencing issue. 
In addition to identifying the specific troubled component,
the context of the issue is provided including exact time of the event, its geographical location, an estimation of the problem severity and the list of the additional components affected by this failure.
It is important to provide the context of the event in order to assist the operator to discover the root cause of the problem and act quickly to restore normal system behavior.

%{\small \begin{verbatim}
%Slack notification:
%Component XXX is having connectivity problems
%Start time: YYYY:MM:DD HH:mm:ss
%Alert time: YYYY:MM:DD HH:mm:ss
%Problem severity: (1-100%)
%Component type:
%List of affected components:
%Location: (e.g., geo location)
%Further details:
%        (e.g., link to a connectivity matrix)
%\end{verbatim}}

\section{CONCLUSIONS} \label{sec:concl}
We have presented an AIOps solution that provides insights useful for the operation of the IBM COS service. 
%Our solution is based on data produced as part of service operation, enriched with auxiliary information in near real time. 
The solution is now in the process of being globally deployed across multiple service offerings in IBM Cloud and will be further refined, optimized, and extended, e.g., to work with more cloud services and for cross-service operations. 
To conclude the paper we share a concise summary of the most important lessons and best practices that can illuminate the path to success for others who pursue similar goals.    

\emph{Know your tools.} There are multiple applicable tools and methods and it is very important to have good understanding of the suitability, efficiency, and compatibility of these methods in the context of a particular operational challenge. 
%As there are no uniformly established best practices for applying data driven analytics for IT operations, maintaining and sharing the taxonomy of tools and methodologies is of ultimate importance.

\emph{Keep it simple.} This timeless wisdom is your best friend when developing AIOps. The appeal of advanced analytics, including machine and deep learning, is so great that many fall in the trap of needless over complication. 
%Some AIOps systems are far more complex and resource hungry than necessary, while simpler, e.g. statistical, analysis can uncover invaluable results, at least for initial problem identification.

\emph{Iterate and refine as you go.} In an ideal world, one assumes sufficient and reliable data sources and specific questions to guide AIOps exploration. In reality, however, the data sources often are not intended for analytical purposes and asking the right questions is a significant challenge. Use an iterative approach whereby insights are gained incrementally, using snapshots, historical data, and calibration.

\emph{Feed and sustain the engagement.} Last but not least, do not underestimate the importance of mutual trust between IT system owners and the AIOps team. System owners are often overwhelmed by their day-to-day load and if not seeing immediately useful results, may disengage and lose interest. Great cross-team collaboration is key to putting in place an AIOps system alongside the production service in a reasonable amount of time.
%Gladly, due to great cross-team collaboration, we've been able to gain valuable results and put in place an AIOps system alongside the production service in a reasonable amount of time.

\addtolength{\textheight}{-12cm}   % This command serves to balance the column lengths
                                  % on the last page of the document manually. It shortens
                                  % the textheight of the last page by a suitable amount.
                                  % This command does not take effect until the next page
                                  % so it should come on the page before the last. Make
                                  % sure that you do not shorten the textheight too much.

%%%%%%%%%%%%%%%%%%%%%%%%%%%%%%%%%%%%%%%%%%%%%%%%%%%%%%%%%%%%%%%%%%%%%%%%%%%%%%%%

%%%%%%%%%%%%%%%%%%%%%%%%%%%%%%%%%%%%%%%%%%%%%%%%%%%%%%%%%%%%%%%%%%%%%%%%%%%%%%%%

%%%%%%%%%%%%%%%%%%%%%%%%%%%%%%%%%%%%%%%%%%%%%%%%%%%%%%%%%%%%%%%%%%%%%%%%%%%%%%%%

%%%%%%%%%%%%%%%%%%%%%%%%%%%%%%%%%%%%%%%%%%%%%%%%%%%%%%%%%%%%%%%%%%%%%%%%%%%%%%%%

\bibliography{bibtex}
\bibliographystyle{IEEEtran}

\end{document}